\documentclass[onecolumn, 12pt]{article}
\usepackage{graphicx, amsmath, amsfonts, multirow, textcomp, lscape, setspace, subfigure, color, footnote, enumitem, centernot, amssymb, tikz, nicefrac, extarrows}
\usepackage{xcolor}
\usepackage[vmargin=2cm,hmargin=2cm]{geometry}
\usepackage{bm,fouriernc} %
\usepackage[T1]{fontenc}
\usepackage[mathcal]{euscript}
\usepackage[round]{natbib}
\usetikzlibrary{calc,arrows.meta,shapes,decorations,patterns,decorations.text,shadows,matrix,arrows,decorations.pathreplacing,calligraphy,fadings,shadings,backgrounds}
\tikzstyle{da}=[-{Latex[length=1mm,angle'=90]},thick]
\tikzstyle{dd}=[{Latex[length=1mm,angle'=90]}-{Latex[length=1mm,angle'=90]},thick]
\tikzstyle{te}=[-{Circle[open]},thick]
\tikzstyle{co}=[-,thick]

\newcommand{\eqtt}[1]{\xlongequal{\text{#1}}}

\newcommand\indep{\protect\mathpalette{\protect\independenT}{\perp}}
\def\independenT#1#2{\mathrel{\rlap{$#1#2$}\mkern2mu{#1#2}}}

\def\Ntid{\tilde{N}}

\title{\textbf{Causal Inference in the Multiverse of Hazard}}
\author{En-Yu Lai and Yen-Tsung Huang\textsuperscript{*}\\[2mm]
\normalsize Institute of Statistical Science, Academia Sinica,\\
\normalsize No.128, Academia Road, Section 2, Nankang, Taipei 11529, Taiwan;\\
\normalsize E-mail: ythuang@stat.sinica.edu.tw}
  
\begin{document}
\date{}
\maketitle


\begin{abstract}
    Hazard serves as a pivotal estimand in both practical applications and methodological frameworks. 
    However, its causal interpretation poses notable challenges, including inherent selection biases and ill-defined populations to be compared between different treatment groups.
    In response, we propose a novel definition of counterfactual hazard within the framework of possible worlds. 
    Instead of \textit{conditioning} on prior survival status as a conditional probability, our new definition involves \textit{intervening} in the prior status, treating it as a marginal probability.
    Using single-world intervention graphs, we demonstrate that the proposed counterfactual hazard is a type of controlled direct effect. 
    Conceptually, intervening in survival status at each time point generates a new possible world, where the proposed hazards across time points represent risks in these hypothetical scenarios, forming a ``multiverse of hazard.'' 
    The cumulative and average counterfactual hazards correspond to the sum and average of risks across this multiverse, respectively, with the actual world's risk lying between the two. 
    This conceptual shift reframes hazards in the actual world as a collection of risks across possible worlds, marking a significant advancement in the causal interpretation of hazards. 
\end{abstract}

\textbf{keyword}
	causal inference; counterfactual process; hazard; multiverse; survival analysis

\section{Introduction}
\label{ss:intro}

Hazard plays a vital role in both applied and methodological research. Traditionally defined as a conditional probability, it represents the likelihood of experiencing an outcome at time $t$ given its absence up to time $t^-$. This conditional perspective holds practical appeal, often surpassing the marginal probability. For instance, amid the COVID-19 pandemic, individuals may be more concerned about their probability of contracting the virus, given their prior health status, than the overall marginal probability since December 2019. Methodologically, hazard stands as a fundamental estimand in classic life table analyses~\citep{reed_merrell, keyfitz_frauenthal} and serves as a crucial intermediary in the Nelson-Aalen \citep{nelson_1972, aalen_1978} and Kaplan-Meier \citep{km} estimators. In regression modeling, hazard-based association measures can also be conveniently obtained by the Cox proportional hazards model \citep{cox} and Aalen's additive hazards model \citep{aalen_89}. 

Despite the practical and methodological value of hazard, concerns regarding its causal interpretation have been widely acknowledged~\citep{greenland_96, hernan_haz,aalen_haz_2015, martinussen, didelez}. Firstly, hazard inherently carries a selection bias. Even in randomized studies, the balance between treatment and placebo groups may be disrupted over time due to differential depletion of susceptibles, particularly if the treatment has a causal effect~\citep{hernan_haz}. Secondly, hazard-based effect estimates compares different populations, which leads to difficulty in attributing causality to a well-defined population, even in a counterfactual framework~\citep{martinussen, didelez}. Lastly, hazard is a non-collapsible quantity, meaning that the marginal hazard may not be recovered by a weighted average of conditional hazards \citep{greenland_96, aalen_haz_2015}. 

To address the above issues, \citet{aalen_haz_2015} defined a causal hazard, described as a ``controlled direct effect'' that bypasses confounders and prior survival status. While this causal hazard addresses the built-in selection bias,  there remains ambiguity regarding the specific target population it pertains to. \citet{martinussen}~proposed an alternative approach by defining a counterfactual hazard as a probability conditional on a principal stratum~\citep{principal_stratum} of survivors who would have survived under any type of treatment. However, the proposed principal stratum depends on time and again, may lead to an ill-defined population during the follow-up. Alternatively, researchers may opt for risk-based approaches where the causal interpretation and target population are more straightforward to comprehend \citep{hernan_haz, didelez}. 

This paper aims to establish a conceptual framework for hazard by introducing a pseudo-population concept across the multiverse \citep{multiverse_quantum, multiverse_phils, multiverse_film}. We propose a counterfactual hazard as a marginal probability (Section~\ref{ss:definition}). We delineate its underlying counterfactual population in the multiverse under a possible world framework and elucidate an interpretation of the proposed hazard as risks in the pseudo-populations of the multiverse (Section~\ref{ss:multiverse}). The proposed counterfactual hazard can further be summed up across the multiverse as a cumulative hazard or averaged as an average hazard. We also show that the risk in the actual world is bounded by the average and cumulative hazards (Section~\ref{ss:bounds}). 

\section{Counterfactual hazards: cCT- and iCP-hazards }
\label{ss:counthaz}

\subsection{Potential outcome framework}
We briefly introduce the potential outcome framework \citep{rubin_78}. 
Here, we denote the independence of variables $A$ and $B$ conditional on $C$ as $A\indep B\mid C$.
Let $Z$ represent the treatment, $Y$ the outcome, and $Y(z)$ the counterfactual outcome given $Z=z$.
The aim is to compare the counterfactual outcomes of different treatments, e.g., the difference or ratio of $Y(z_a)$ and $Y(z_b)$. Two assumptions are required to identify $Y(z)$: \textit{Exchangeability} states that the counterfactual outcome is independent of the actual treatment, i.e., $Y(z)\indep Z$; \textit{consistency} states that the counterfactual outcome is identical to the actual outcome when the intervention of treatment is the same as the actual treatment, i.e., $E(Y(z)|Z=z)=E(Y|Z=z)$. 
Under these assumptions, the counterfactual outcome can be identified by 
\[E(Y(z))\eqtt{exchangeability} E(Y(z)\mid Z=z)\eqtt{consistency} E(Y\mid Z=z).\]
Note that we need to consider sufficient confounders to achieve exchangeability in practice, i.e., conditional on the covariates $X$, $Y(z)\indep Z\mid X$. It follows that $E(Y(z)\mid X)=E(Y(z)\mid Z=z,X)=E(Y\mid Z=z,X)$ again, by exchangeability and consistency. When mediators exist between the exposure and the outcome, ensuring exchangeability among the exposure, mediators, and outcome requires assumptions known as \textit{no unmeasured confounding} or \textit{sequential ignorability}. 

\subsection{Definitions and assumptions}
\label{ss:definition}

We let $Z_i$, $T_i$, and $C_i$, respectively, denote the treatment, survival time, and censoring time for subject $i$ where $i=1,...,m$ and $m$ is the sample size and assume that $T_i$ and $C_i$ share the same origin and are independent conditional on $Z_i$ and $X_i$, i.e., noninformative censoring, where $X_i$ are measured covariates. We also define the following processes: $\Ntid_i(t)=I(T_i\leq t)$ denotes the underlying event process where $I(\cdot)$ is an indicator function; $N_i(t)=I(T_i\leq t, T_i\leq C_i)$ denotes the observed event process; $Y_i(t)=I(T_i\geq t, C_i\geq t)$ denotes the at-risk process. We further introduce notations based on counterfactuals. $T_i(z)$ denotes counterfactual survival time for subject $i$ whose $Z_i$ had been set to $z$, and $\Ntid_i(t; z, n(t^-))$ denotes a counterfactual event process \citep{huang_semicomp} at time $t$ had $Z_i$ and $\Ntid_i(t^-)$ been set to $z$ and $n(t^-)$, respectively, where $n(t^-)\in\{0,1\}$. The intervention $\Ntid_i(t^-)=n(t^-)$ creates a new possible world at each $t$ that will be illustrated in Section~\ref{ss:multiverse}. 

We introduce two counterfactual hazards, with one based on the counterfactual survival time $T(z)$ (cCT) and the other using the counterfactual event process (iCP):
\begin{align}
d\Lambda_{cCT}(t\mid z)&:=P\{T(z)\in[t, t+dt)\mid T(z)\geq t\}\label{eq:cCT_def}\\
d\Lambda_{iCP}(t\mid z)&:=P\{\Ntid(t; z, n(t^-)=0)=1\}. \label{eq:iCP_def}
\end{align}
The cCT (conditioning counterfactual survival time) defines the hazard by conditioning on the event of $T(z)\geq t$ whereas the iCP (interventional counterfactual process) does so by an intervention setting $\Ntid(t^-)=n(t^-)=0$. 
Note that $n(t^-)$ represents an intervention $\in\{0,1\}$ depending on time; $P\{\Ntid(t; z, n(t^-)=0)=1\}$ can also be expressed using the \textit{do} operator by \citet{pearl_book}: $P\{\Ntid(t)=1\mid do(Z=z, \Ntid(t^-)=0)\}$, which is closely related to the causal hazard by \citet{aalen_haz_2015}. The difference is that \citet{aalen_haz_2015} intervenes in a survival function, and the iCP intervenes in a Bernoulli variable $\Ntid(t^-)$. 

We further denote the two hazards conditional on covariates $X=x$: 
\begin{align*}
d\Lambda_{cCT}(t\mid z, x)&:=P\{T(z)\in[t, t+dt)\mid T(z)\geq t, X=x\}\\
d\Lambda_{iCP}(t\mid z, x)&:=P\{\Ntid(t; z, n(t^-)=0)=1\mid X=x\}. 
\end{align*}
Under the assumption of $T(z)\indep Z\mid X$, we have the following identification formula for $d\Lambda_{cCT}(t \mid z, x)$:
\begin{align*}
d\Lambda_{cCT}(t\mid z, x):=&P\{T(z)\in[t, t+dt)\mid T(z)\geq t, X=x\}\\
    =&P(T(z)\in[t, t+dt)\mid T(z)\geq t, Z=z, X=x)&&\text{\textit{exchangeability}}\\
    =&P(T\in[t, t+dt)\mid T\geq t, Z=z, X=x).&&\text{\textit{consistency}}
\end{align*}
Under the assumption of $\Ntid(t; z, n(t^-))\indep (Z, \Ntid(t^-))\mid X$, we have
\begin{align*}
d\Lambda_{iCP}(t\mid z, x):=&P\{\Ntid(t; z, n(t^-)=0)=1\mid X=x\}\\
    =&P(\Ntid(t; z, n(t^-)=0)=1\mid \Ntid(t^-)=0, Z=z, X=x)&&\text{\textit{exchangeability}}\\
    =&P(d\Ntid(t)=1\mid \Ntid(t^-)=0, Z=z, X=x).&&\text{\textit{consistency}}
\end{align*}
where $d\Ntid(t)=\Ntid(t)-\Ntid(t^-)$, and the formula is equivalent to that for $d\Lambda_{cCT}(t\mid z, x)$.
By the definition of $d\Lambda_{iCP}(t\mid z)$, it follows that  
\begin{align*}
d\Lambda_{iCP}(t\mid z)&=\sum_xd\Lambda_{iCP}(t\mid z, x)P(X=x)\\
&=\sum_xP(d\Ntid(t)=1\mid \Ntid(t^-)=0, Z=z, X=x)P(X=x),
\end{align*}
which, however, is different from that for $d\Lambda_{cCT}(t\mid z)$ (derived under $T(z)\indep Z\mid X$):
\begin{align*}
d\Lambda_{cCT}(t\mid z)&=\frac{\sum_xP(T\in[t, t+dt)\mid Z=z, X=x)P(X=x)}{\sum_xP(T\geq t\mid Z=z, X=x)P(X=x)}\\
&\neq \sum_xd\Lambda_{cCT}(t\mid z, x)P(X=x),
\end{align*}
due to the non-collapsibility of a conditional probability. 
Under noninformative censoring, we can construct the following estimators
\begin{align}
d\hat{\Lambda}_{cCT}(t\mid z)&=\frac{\sum_{x}\sum_{i=1}^mdN_i(t)I(X_i=x, Z_i=z)\sum_{i=1}^{m}I(X_i=x)}{\sum_{x}\sum_{i=1}^mY_i(t)I(X_i=x, Z_i=z)\sum_{i=1}^{m}I(X_i=x)}\label{eq:cCT}\\
d\hat{\Lambda}_{iCP}(t\mid z)&=\sum_{x}\frac{\sum_{i=1}^mdN_i(t)I(X_i=x, Z_i=z)}{\sum_{i=1}^mY_i(t)I(X_i=x, Z_i=z)}\frac{1}{m}\sum_{i=1}^mI(X_i=x).\label{eq:iCP}
\end{align}\label{eq:est}

\subsection{Graphical illustration}
\label{ss:dag}

\begin{figure}
    \centering
    \subfigure[iCP as a controlled direct effect]{\scalebox{1}{\begin{tikzpicture}
\def\bb{teal!80!blue}
\def\gg{blue!20!red}

\node (Z) {$\textcolor{\bb}{z}$};
\node (N1) at (1.5,0) {$\Tilde{N}(t_1;z)$};
\node (N2) at (3.8,0) {$\textcolor{\gg}{n(t_2)=0}$};
\node (N3) at (7,0) {$\Tilde{N}(t_3;\textcolor{\bb}{z};\textcolor{\gg}{n(t_2)=0})$};

\draw[da] (Z) -- (N1);
\draw[da] (N2) edge[\gg] (N3);
\draw[da] (Z) edge[\bb,bend left=60] ($(N3)+(-.5,.2)$);

\end{tikzpicture}}}
    \subfigure[exposure-outcome confounder needs to be adjusted]{\scalebox{1}{\begin{tikzpicture}
\def\bb{teal!80!blue}
\def\gg{blue!20!red}
\def\yy{red!40!yellow}

\node (Z) at (-.3,0) {$Z|\textcolor{\bb}{z}$};
\node (N1) at (1.5,0) {$\Tilde{N}(t_1;z)$};
\node (N2) at (3.8,0) {$\textcolor{\gg}{n(t_2)=0}$};
\node (N3) at (7,0) {$\Tilde{N}(t_3;\textcolor{\bb}{z};\textcolor{\gg}{n(t_2)=0})$};
\node (X) at ($(N2)+(0,-1.8)$) {$\textcolor{\yy}{X}$};

\draw[da] (Z) -- (N1);
\draw[da,\gg] (N2) -- (N3);
\draw ($(Z)+(.3,.2)$) edge[da,\bb,bend left=55] ($(N3)+(-.5,.2)$);
\draw[da] (X) edge[\yy] ($(Z)+(-.2,-.3)$);
\draw[da] (X) edge[\yy] (N1);
\draw[da] (X) edge[\yy] (N3);

\end{tikzpicture}}}
    \subfigure[outcome-outcome confounder needs to be adjusted]{\scalebox{1}{\begin{tikzpicture}
\def\bb{teal!80!blue}
\def\gg{blue!20!red}
\def\yy{red!40!yellow}

\node (Z) at (-.3,0) {$Z|\textcolor{\bb}{z}$};
\node (N1) at (1.5,0) {$\Tilde{N}(t_1;z)$};
\node (N2) at (4,0) {$\Tilde{N}(t_2)|\textcolor{\gg}{n(t_2)=0}$};
\node (N3) at (8,0) {$\Tilde{N}(t_3;\textcolor{\bb}{z};\textcolor{\gg}{n(t_2)=0})$};
\node (X) at ($(N2)+(0,-1.8)$) {$\textcolor{\yy}{X}$};
\node (e1) at ($(N1)+(0,-1)$){$\epsilon_1$};
\node (e3) at ($(N3)+(0,-1)$){$\epsilon_3$};

\draw[da] (Z) -- (N1);
\draw[da,\gg] (N2) -- (N3);
\draw ($(Z)+(.3,.2)$) edge[da,\bb,bend left=60] ($(N3)+(-.5,.2)$);
\draw ($(Z)+(0,.3)$) edge[da,bend left=55,gray] ($(N2)+(-.7,.2)$);
\draw[da,\yy] (X) -- ($(N2)+(-.7,-.3)$);
\draw[da,\yy] (X) -- (N1);
\draw[da,\yy] (X) -- (N3);
\draw[<-] (N1) -- (e1);
\begin{scope}[transform canvas={xshift =-8mm}]
    \node (e2) at ($(N2)+(0,-1)$){$\epsilon_2$};
    \draw[<-] (N2) -- (e2);
\end{scope}
\draw[<-] (N3) -- (e3);

\end{tikzpicture}}}
    \caption{Graphical illustrations of the iCP hazard. }
    \label{ff:dag}
\end{figure}
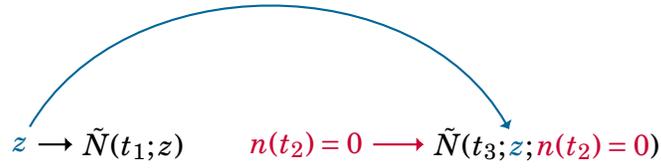
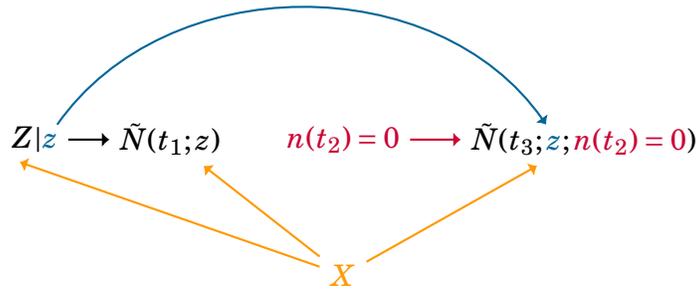
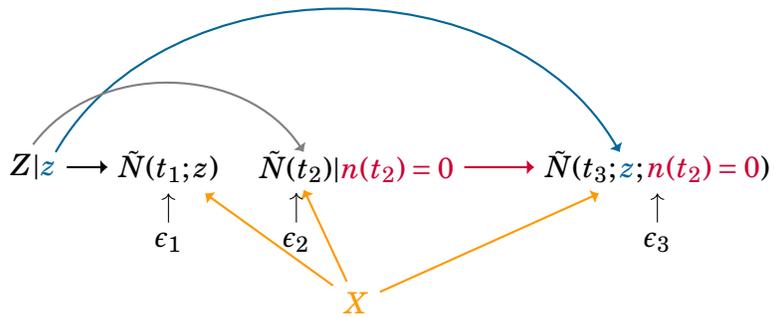

Given the complexity of depicting counterfactual survival time in a time-dependent manner on causal diagrams, our focus here is on the iCP hazard, where the process can be discretized into a series of Bernoulli variables, such as $\Ntid(t_1)$, $\Ntid(t_2)$, and so forth. We illustrate the iCP hazard using single-world intervention graphs (SWIGs)~\citep{swig}. Comparing a counterfactual hazard between different values of $Z$ may be viewed as a direct causal effect of $Z$ on the outcome not through the earlier outcomes. This is achieved by the iCP via setting $\Ntid(t^-)=n(t^-)=0$ (depicted by the blue arrow in Figure~\ref{ff:dag}(a)). Adjustments for the exposure-outcome confounder are necessary to ensure $\Ntid(t; z, n(t^-))\indep Z\mid X$ (Figure~\ref{ff:dag}(b)), and adjustments for the outcome-outcome confounder (Figure~\ref{ff:dag}(c)) are needed to ensure $\Ntid(t; z, n(t^-))\indep \Ntid(t^-)\mid X$. Similar to identifying controlled direct effects in mediation analyses \citep{vanderweele_2009, imai_2009}, the iCP hazard is identifiable in the presence of treatment-induced outcome-outcome confounder that is construed as a type of outcome-outcome confounders. 

The iCP hazard is also similar to the survivor average causal effect (SACE)~\citep{sace_rubin} comparing instantaneous risks of $Z=1$ versus $Z=0$ in the principal stratum $\{T(z=0)\geq t, T(z=1)\geq t\}$ \citep{martinussen}. 
Identifying the SACE relies on a strong assumption, $T(z=1)\indep T(z=0)\mid Z$, which may be fulfilled by controlling for all variability in $T$.
As depicted in Figure~\ref{ff:dag}(c), the iCP hazard still allows time-specific variability for $\Ntid(t)$ (i.e., $\epsilon_1$, $\epsilon_2$, and $\epsilon_3$). 
This highlights another advantage of the counterfactual counting process over counterfactual survival time: the ability to explicitly represent the causal mechanism in a time-dependent setting.

\subsection{A data example using the REVEAL cohort}

We present a real-world application to showcase the advantages of iCP hazards. The data were obtained from a community-based cohort in Taiwan as part of the Risk Evaluation of Viral Load Elevation and Associated Liver (REVEAL) study~\citep{chen2006risk}.
In this study, the exposure of interest is the result of the hepatitis B surface antigen (HBsAg) test. A positive result indicates that the patient carries the HBV virus. The survival outcome is defined as the time to hepatocellular carcinoma.
Additionally, we include covariates such as age, gender, smoking and drinking statuses, and alanine transaminase levels to account for potential confounding factors.

\begin{figure}
    \centering
    \includegraphics[width=.8\linewidth]{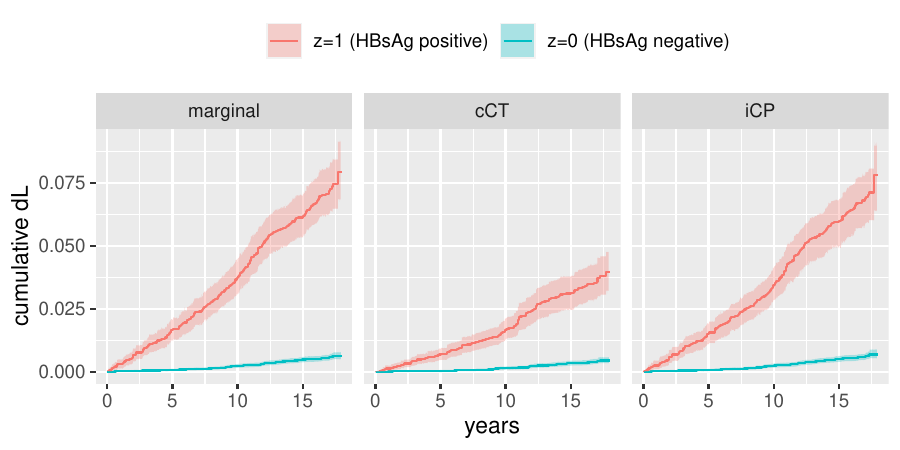}
    \caption{Cumulative marginal, cCT, and iCP hazards using the REVEAL cohort.}
    \label{fig:REVEAL}
\end{figure}

In Figure~\ref{fig:REVEAL}, we compare the cumulative iCP, cCT, and marginal hazard curves. The marginal curve is derived from a naive Nelson-Aalen estimator without considering any covariates.
If covariate adjustment is not considered in the model, cCT and iCP will yield identical results numerically. However, the nature of the two estimands differs, and this dissimilarity becomes evident when integrating stratified hazards that are conditional on covariates.
As stated in Equation~\eqref{eq:cCT_def}, the cCT hazard is defined as a conditional probability. Therefore, the appropriate method for integrating covariates involves separately integrating the numerator and denominator to obtain a hazard of a standardized population, as shown in Equation~\eqref{eq:cCT}.
In contrast, the iCP hazard in Equation~\eqref{eq:iCP_def} is defined as a marginal probability. Thus, integrating covariates simply entails the weighted average of hazards, as demonstrated in Equation~\eqref{eq:iCP}.

To achieve such causal interpretability, we assume the independence of outcome status across different time points conditional on sufficient covariates, i.e., $\Ntid(t; z, n(t^-))\indep (Z, \Ntid(t^-))\mid X$. This assumption implies that the population at time $t$ and $t^-$ should be recognized as two identical but independent worlds, akin to the multiverse interpretation we introduce later. Consequently, any difference between these two worlds can be attributed solely to the intervened variable $Z$.

It's crucial to highlight that using cCT or classical hazards (which do not involve counterfactuals but still rely on conditioning on the time variable) can suffer from non-collapsibility. As depicted in Figure~\ref{fig:REVEAL}, the scale of the cCT hazard appears much smaller than that of the marginal hazard since it averages the effect within the population. Consequently, distinguishing whether the difference originates from confounding effects or non-collapsibility can be challenging.
In contrast, iCP hazards are marginal probabilities and free of non-collapsibility. Therefore, any difference between iCP and marginal hazards can be readily attributed to confounding effects.

Biologically speaking, the HBsAg-positive population for calculating the cCT hazard became healthier over time as individuals can be affected but not yet diseased. However, the population for calculating the iCP hazard remained the same over time. Consequently, the cCT hazard will underestimate the risk of hepatocellular carcinoma. A toy example demonstrates selection bias is presented in Figure~\ref{fig:hz} (first panel), where the control population became healthier due to the beneficial effect of treatment.

\section{Multiverse of hazards}
\label{ss:multiverse}
In this paper, we leverage elements from possible-world semantics~\citep{Lewis1973-LEWC-2,sep-possible-worlds}, a classical method used to determine the truth value of counterfactual conditionals. We adapt these concepts to develop a customized semantics for counterfactual hazards, which we refer to as the \textit{multiverse semantic}.
In our framework, the \textit{actual world} represents the reality that we observe events and collect data from. A \textit{possible world} is a virtual counterpart to the actual world, where we can introduce artificial settings to calculate statistics with causal interpretation.
Within this framework, an \textit{agent} refers to a statistician who observes events and computes statistics through these worlds.

\subsection{Direct interpretation}

As illustrated in Figure~\ref{fig:hz} (first panel), participants surviving in the control group tend to be inherently healthier than those in the treatment group over time, assuming treatment benefits.
The core concept behind the iCP-hazard is to compute hazards from possible worlds where control and treatment groups remain comparable. This world is achieved by intervening with $n(t^-)=0$, effectively reviving all deceased individuals at $t^-$ in the actual world.
In this possible world, those who die at $t$ in the actual world will also perish, while those revived at $t$ will experience their potential outcomes that they would not have had in the actual world.
Using this semantics, the agent calculates hazard at each time point $t$ by simulating a scenario where all deceased individuals are revived by $t^-$ in the actual world. This approach is depicted at the individual level in Figure~\ref{fig:hz} (second panel).

\subsection{Multiverse interpretation}
As depicted in Figure~\ref{fig:chz} (upper panel), the actual risk evaluates the same population and total deaths within a specific time frame. However, the cumulative iCP-hazard considers different pseudo-populations across various time slices, where the sample size fluctuates throughout the follow-up due to the die-and-revive intervention.
To circumvent the need for reviving interventions, we propose an alternative interpretation for the iCP-hazard: the multiverse interpretation. This approach provides a clearer conceptualization of the pseudo-population underlying the cumulative iCP-hazard.

Instead of individually reviving the deceased at each event time (the direct interpretation), we establish a series of counterfactual scenarios from the outset, identical to the actual world at $t_0$. In each of these possible worlds, individuals can only perish at a predetermined time point, as depicted in Figure~\ref{fig:hz} (third panel). This ensemble of possible worlds is termed the \textit{multiverse of hazard}.
Under this framework, the $j$-th event time, denoted as $t_j$, in the actual world corresponds to the $j$-th possible world, created by setting $n(t_j^-)=0$, allowing the agent to compute the hazard in each scenario. Similar to the direct interpretation, individuals who perish at $t_j$ in the actual world will also perish in the $j$-th possible world. Those who die before $t_j$ in the actual world receive their potential outcomes in the $j$-th possible world, outcomes not observed had they died earlier in the actual world.
Since no deaths occur before $t_j$ in the $j$-th possible world, the comparability of control and treatment groups at $t^-$ mirrors that at baseline. Consequently, the hazard from each possible world inherits a valid causal interpretation, provided the two groups are randomized at baseline.
Moreover, since no deaths occur after $t_j$, the hazard represents the risk in each possible world. The cumulative hazard using this semantics can be interpreted as the summation of cross-world hazards (or equivalently, the total risk of the multiverse), which can be compared to the actual risk, as illustrated in the boxed panel of Figure~\ref{fig:chz} and detailed in Section~\ref{ss:bounds}.

In summary, Figure~\ref{fig:chz} (boxed panel) illustrates three distinct routes for the agent.
The black path signifies the conventional hazard assessment, where the agent remains in the actual world throughout the follow-up, observing cross-time survival statuses and calculating both classical hazards and actual risk. The gray path represents the direct interpretation of the iCP-hazard. Here, the agent traverses between actual and possible worlds, observing cross-world survival statuses and computing the counterfactual hazard.
On the other hand, the green, red, and yellow paths depict the multiverse interpretation of the iCP hazard. In this scenario, the agent journeys within each possible world, tracking survival statuses over time and computing the counterfactual hazard as a risk. Conceptually, this statistician agent takes on a super-hero task similar to that of Dr. Strange in Marvel \citep{multiverse_film}. 

\begin{figure}
    \centering
    \scalebox{.46}{\begin{tikzpicture}

\def\ww{1}
\def\hh{.5}
\def\rr{120}
\newcommand{\colorD}{gray!30}
\newcommand{\colorR}{yellow!80!black}
\newcommand{\colorTone}{blue!50!green!40!black}
\newcommand{\colorTtwo}{red!40!black}
\tikzstyle{bb}=[circle,draw,thick]
\tikzstyle{dd}=[pattern=crosshatch,pattern color=\colorD,draw=\colorD]
\tikzstyle{rr}=[draw=\colorR,line width=1mm]
\tikzstyle{rd}=[pattern=crosshatch,pattern color=\colorD,draw=\colorR,line width=1mm]

\newcommand{\CC}[9]{
\node[bb,#9] (A1) {$A_{#1}$}; 
\pgfmathtruncatemacro\result{#1+1};
\node[bb,#9] (A2) at (1*\ww,.5*\hh) {$A_{\result}$}; 
\pgfmathtruncatemacro\result{#1+2};
\node[bb,#9] (A3) at (.8*\ww,-1.5*\hh) {$A_{\result}$}; 
\node[bb,#9] (B1) at ($(A1)+(2.5*\ww,\hh)$) {$B_{#2}$}; 
\pgfmathtruncatemacro\result{#2+1};
\node[bb,#9] (B2) at ($(A2)+(2.5*\ww,\hh)$) {$B_{\result}$}; 
\pgfmathtruncatemacro\result{#2+2};
\node[bb,#9,#4] (B3) at ($(A3)+(2.5*\ww,\hh)$) {$B_{\result}$}; 
\pgfmathtruncatemacro\result{#2+3};
\node[bb,#9,#5] (B4) at ($(B2)+(.8*\ww,-1.5*\hh)$) {$B_{\result}$}; 
\node[bb,#9,#6] (C1) at ($(A1)+(2*\ww,-4*\hh)$) {$C_{#3}$}; 
\pgfmathtruncatemacro\result{#3+1};
\node[bb,#9,#7] (C2) at ($(A2)+(2*\ww,-4*\hh)$) {$C_{\result}$}; 
\pgfmathtruncatemacro\result{#3+2};
\node[bb,#9,#8] (C3) at ($(A3)+(2*\ww,-4*\hh)$) {$C_{\result}$}; 
}

\node[bb] (A) at (21.5,5) {\scriptsize $A$};
\node[anchor=west] (AM) at ($(A)+(.4*\ww,0)$) {\footnotesize\textit{healthy people}};
\node[bb] (B) at ($(A)+(0,-1.6*\hh)$) {\scriptsize $B$};
\node[anchor=west] (BM) at ($(B)+(.4*\ww,0)$) {\footnotesize\textit{average people}};
\node[bb] (C) at ($(A)+(0,-3.2*\hh)$) {\scriptsize $C$};
\node[anchor=west] (CM) at ($(C)+(.4*\ww,0)$) {\footnotesize\textit{unhealthy people}};
\node[bb] (L) at ($(A)+(0,-4.8*\hh)$) {\scriptsize\textcolor{white}{A}};
\node[anchor=west] (LM) at ($(L)+(.4*\ww,0)$) {\footnotesize\textit{alive}};
\node[bb,dd] (D) at ($(A)+(0,-6.4*\hh)$) {\scriptsize\textcolor{white}{A}};
\node[anchor=west] (DM) at ($(D)+(.4*\ww,0)$) {\footnotesize\textit{actually dead}};
\node[bb,rr] (RR) at ($(A)+(0,-8*\hh)$) {\scriptsize\textcolor{white}{A}};
\node[anchor=west] (RRM) at ($(RR)+(.4*\ww,0)$) {\footnotesize\textit{counterfactually alive}};
\node[bb,rd] (RD) at ($(A)+(0,-9.6*\hh)$) {\scriptsize\textcolor{white}{A}};
\node[anchor=west] (RDM) at ($(RD)+(.4*\ww,0)$) {\footnotesize\textit{counterfactually dead}};
\draw[rounded corners,gray!10,line width=2mm] ($(A)+(-.6,.6)$) rectangle ($(A)+(4.2,-5.4)$);
\draw[->,thick] (-.5,3) -- (19.5,3) node [pos=0.95,yshift=-5mm] {time};

\CC{1}{1}{1}{}{}{}{}{}{}
\draw ($(B1)+(0,2.2)$) -- ($(B1)+(0,2.8)$) node [yshift=-1cm] {$t_0$} node [yshift=4mm] {\textit{randomization}};
\begin{scope}[transform canvas={xshift = 7cm}]
\CC{1}{1}{1}{}{dd}{}{}{dd}{}
\draw ($(B1)+(0,2.2)$) -- ($(B1)+(0,2.8)$) node [yshift=-1cm] {$t_1$} node [yshift=11mm,text width=51mm,align=flush center] {calculating hazard at $t_1$ is of well causal interpretation because population at $t_1^{-}$ are comparable};
\end{scope}
\begin{scope}[transform canvas={xshift = 14cm}]
\CC{1}{1}{1}{dd}{dd}{dd}{dd}{dd}{}
\draw[rounded corners,dashed] (-.8,1.5) rectangle (5.1*\ww,-7*\hh);
\node[text width=22mm,align=flush center] (CTT) at (.3,-5.5*\hh) {$\nicefrac{\text{healthy}}{\text{all}}$ $=60\%$};
\draw ($(B1)+(0,2.2)$) -- ($(B1)+(0,2.8)$) node [yshift=-1cm] {$t_2$} node [yshift=11mm,text width=51mm,align=flush center] {calculating hazard at $t_2$ is of ill causal interpretation because population at $t_2^{-}$ are not comparable};
\end{scope}
\begin{scope}[transform canvas={yshift = -5.2cm}]
\CC{4}{5}{4}{}{}{}{}{}{}
\end{scope}
\begin{scope}[transform canvas={yshift = -5.2cm,xshift = 7cm}]
\CC{4}{5}{4}{}{}{}{dd}{}{}
\end{scope}
\begin{scope}[transform canvas={yshift = -5.2cm,xshift = 14cm}]
\CC{4}{5}{4}{}{dd}{}{dd}{}{}
\draw[rounded corners,dashed] (-.8,1.5) rectangle (5.1*\ww,-7*\hh);
\node[text width=22mm,align=flush center] (CTT) at (.3,-5.5*\hh) {$\nicefrac{\text{healthy}}{\text{all}}$ $=37.5\%$};
\end{scope}

\node (CT) at  ($(A3)+(-3,-.5)$) {\textsc{Control}};
\node (TM) at ($(CT)+(0,-5.2)$) {\textsc{Treatment}};
\draw [thick,decorate,decoration={brace,amplitude=15pt}] ($(CT)+(21.3,1)$) -- ($(TM)+(21.3,-1)$) node [pos=.5,xshift=30mm,text width=45mm,align=flush center] {people remaining alive in the control group are healthier than those in the treatment group};
\draw[thick] (-4,-9.5) -- (25.5,-9.5) node [pos=0.06,yshift=5mm] {\large\textbf{actual world}} node [pos=0.23,yshift=-5mm] {\large\textbf{possible world: direct interpretation of the iCP hazard}};

\begin{scope}[transform canvas={yshift = -12cm}]

\begin{scope}[transform canvas={xshift = 7cm}]
\CC{1}{1}{1}{}{rr}{}{}{rr}{}
\draw[rounded corners,dashed] (-.8,1.5) rectangle (5.1*\ww,-7*\hh);
\end{scope}
\begin{scope}[transform canvas={xshift = 14cm}]
\CC{1}{1}{1}{dd}{rr}{dd}{dd}{rd}{}
\node[text width=46mm,align=flush center] (RR) at (8,0) {$B_4$ revives at $t_2^-$ and is counterfactually alive at $t_2$ since the B-population are mostly alive};
\draw[->] (B4) -- (RR);
\node[text width=22mm,align=flush center] (DD) at (0,-2.5) {$B_3$, $C_1$ and $C_2$ die at $t_2$};
\draw[->] (B3) -- (DD);
\draw[->] (C1) -- (DD);
\draw[->] (C2) edge[bend right=35] (DD);
\node[text width=46mm,align=flush center] (RD) at (7,-3) {$C_4$ revives at $t_2^-$ and is counterfactually dead at $t_2$ since the C-population are mostly dead};
\draw[->] (C3) -- (RD);
\end{scope}
\begin{scope}[transform canvas={yshift = -5.2cm,xshift = 7cm}]
\CC{4}{5}{4}{}{}{}{rr}{}{}
\draw[rounded corners,dashed] (-.8,1.5) rectangle (5.1*\ww,-7*\hh);
\end{scope}
\begin{scope}[transform canvas={yshift = -5.2cm,xshift = 14cm}]
\CC{4}{5}{4}{}{dd}{}{rr}{}{}
\node (DD) at (7,0) {$B_8$ dies at $t_2$};
\draw[->] (B4) -- (DD);
\node[text width=46mm,align=flush center] (RR) at (7,-2) {$C_5$ revives at $t_2^-$ and is counterfactually alive at $t_2$ since the C-population are mostly alive};
\draw[->] (C2) -- (RR);
\end{scope}
\node (CT) at ($(A3)+(-3,-.5)$) {\textsc{Control}};
\node (TM) at ($(CT)+(0,-5.2)$) {\textsc{Treatment}};
\draw [thick,decorate,decoration={brace,amplitude=15pt,mirror}] ($(CT)+(8.4,1)$) -- ($(TM)+(8.4,-1)$) node [pos=.5,xshift=-25mm,text width=40mm,align=flush center] {counterfactually randomized at $t_2^{-}$ (by~reviving $B_4$, $C_3$~and~$C_5$)};

\end{scope}
\draw[thick] (-4,-21.2) -- (25.5,-21.2) node [pos=0.25,yshift=-5mm] {\large\textbf{possible worlds: multiverse interpretation of the iCP hazard}};

\begin{scope}[transform canvas={yshift = -24cm}]

\CC{1}{1}{1}{}{}{}{}{}{\colorTone}

\begin{scope}[transform canvas={xshift = 7cm}]
\CC{1}{1}{1}{}{dd}{}{}{dd}{\colorTone}
\end{scope}
\begin{scope}[transform canvas={xshift = 14cm}]
\CC{1}{1}{1}{}{dd}{}{}{dd}{\colorTone}
\end{scope}
\begin{scope}[transform canvas={yshift = -5.2cm}]
\CC{4}{5}{4}{}{}{}{}{}{\colorTone}
\end{scope}
\begin{scope}[transform canvas={yshift = -5.2cm,xshift = 7cm}]
\CC{4}{5}{4}{}{}{}{dd}{}{\colorTone}
\node[text width=22mm,align=flush center] (CTT) at (.3,-5.5*\hh) {chance to die at $t_1$};
\end{scope}
\begin{scope}[transform canvas={yshift = -5.2cm,xshift = 14cm}]
\CC{4}{5}{4}{}{}{}{dd}{}{\colorTone}
\end{scope}

\node[\colorTone] (CT) at  ($(A3)+(-3,-.5)$) {\textsc{Control}};
\node[\colorTone] (TM) at ($(CT)+(0,-5.2)$) {\textsc{Treatment}};
\draw[rounded corners] (6.1,1.5) rectangle (12.2,-8.7);
\draw[rounded corners,blue!50!green!30,line width=1mm] ($(CT)+(-1.7,3)$) rectangle ($(TM)+(21.8,-2.5)$);
\draw[->,blue!50!green!30,line width=1mm] ($(TM)+(21.8,7)$) -- ($(TM)+(23,7)$);
\node[text width=38mm,align=flush center,\colorTone,anchor=west] (W1) at ($(TM)+(23,7)$) {a possible world where people only die at $t_1$};
\draw[->] (12.2,-3.6) -- (20.8,-3.6);
\node[text width=38mm,align=flush center,anchor=west] at (21,-3.6) {people die at $t_j$ in the actual world will also die in the $j$-th possible world};
\end{scope}

\begin{scope}[transform canvas={yshift = -35cm}]

\CC{1}{1}{1}{}{}{}{}{}{\colorTtwo}

\begin{scope}[transform canvas={xshift = 7cm}]
\CC{1}{1}{1}{}{}{}{}{}{\colorTtwo}
\end{scope}
\begin{scope}[transform canvas={xshift = 14cm}]
\CC{1}{1}{1}{dd}{rr}{dd}{dd}{rd}{\colorTtwo}
\end{scope}
\begin{scope}[transform canvas={yshift = -5.2cm}]
\CC{4}{5}{4}{}{}{}{}{}{\colorTtwo}
\end{scope}
\begin{scope}[transform canvas={yshift = -5.2cm,xshift = 7cm}]
\CC{4}{5}{4}{}{}{}{}{}{\colorTtwo}
\end{scope}
\begin{scope}[transform canvas={yshift = -5.2cm,xshift = 14cm}]
\CC{4}{5}{4}{}{dd}{}{rr}{}{\colorTtwo}
\node[text width=22mm,align=flush center] (CTT) at (.3,-5.5*\hh) {chance to die at $t_2$};
\end{scope}

\node[\colorTtwo] (CT) at  ($(A3)+(-3,-.5)$) {\textsc{Control}};
\node[\colorTtwo] (TM) at ($(CT)+(0,-5.2)$) {\textsc{Treatment}};
\draw[rounded corners] (13,1.5) rectangle (19.1,-8.7);
\draw[rounded corners,red!30,line width=1mm] ($(CT)+(-1.7,3)$) rectangle ($(TM)+(21.8,-2.5)$);
\draw[->,red!30,line width=1mm] ($(TM)+(21.8,7)$) -- ($(TM)+(23,7)$);
\node[text width=38mm,align=flush center,\colorTtwo,anchor=west] (W1) at ($(TM)+(23,7)$) {a possible world where people only die at $t_2$};
\draw (19.1,-2.6) -- (20,-2.6);
\draw (20,-2.6) -- (20,7.4);
\draw[->] (19.1,-3.6) -- (20.8,-3.6);
\node[text width=42mm,align=flush center,anchor=west] at (21,-3.6) {people die before $t_j$ in the actual world will present their potential outcomes in the $j$-th possible world};
\node[text width=13mm,align=flush center,draw,rounded corners] at (21.5,-5.7) {total death};
\node[scale=1.2] at (20.4,-7) {$=$};
\node[text width=13mm,align=flush center,draw,rounded corners] at (21.5,-7) {actual death};
\node[scale=1.2] at (22.5,-7) {$+$};
\node[text width=27mm,align=flush center,draw,rounded corners] at (24.2,-7) {counterfactual death};

\end{scope}
\node (BB) at (0,-44) {};
\end{tikzpicture}}
    \caption{Multiverse interpretation of the iCP hazard.}
    \label{fig:hz}
\end{figure}
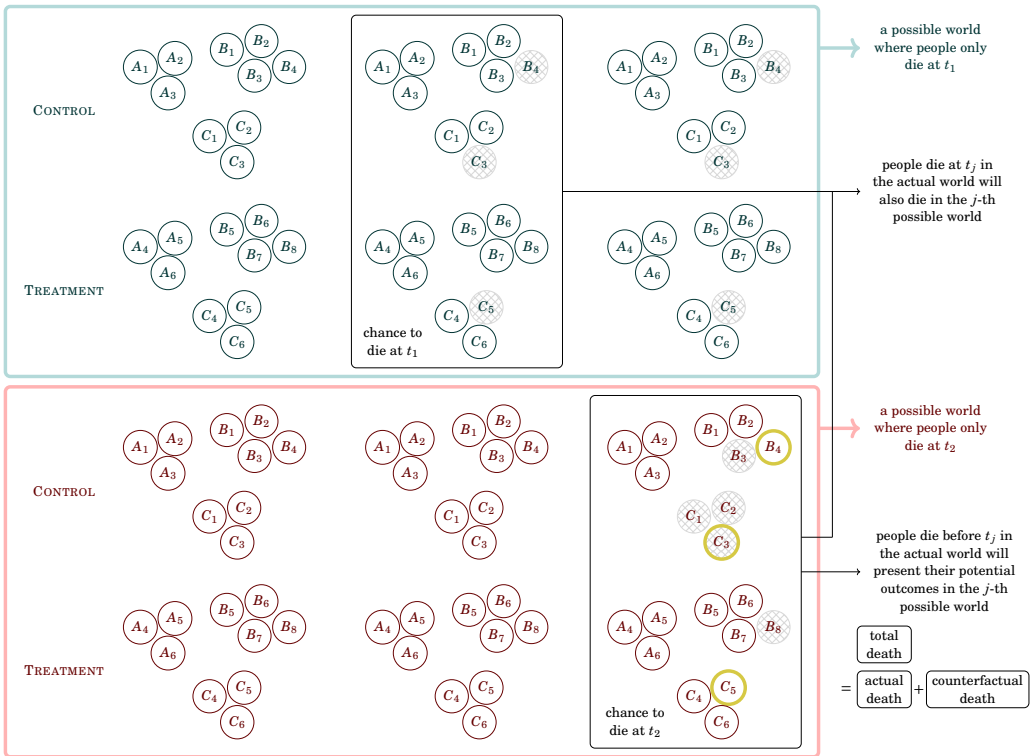

\begin{figure}
    \centering
    \scalebox{.7}{\input{population.tikz}}
    \caption{Multiverse interpretation of the iCP cumulative hazard.}
    \label{fig:chz}
\end{figure}

\section{Relationships of the actual risk and (cumulative and average) iCP-hazard}
\label{ss:bounds}

We omit the treatment $Z$ for succinct notations and consider $J$ discrete event time points: $t_1<\ldots<t_J<\tau$, where $\tau$ marks the study's end. The \textit{cumulative counterfactual hazard}, denoted $\Lambda_{iCP}(\tau)$, sums the ratio of total deaths $d_j$ (including actual and potential deaths) in the $j$-th possible world over the sample size $m$:
\begin{align*}
\Lambda_{iCP}(\tau):=\sum_{j=1}^{J}\frac{d_j}{m},
\end{align*}
Its estimation $\hat{\Lambda}{iCP}(\tau)$ aggregates the counterfactual hazard estimates $\hat{\Lambda}{iCP}(t_j)$ across the event times $t_j$: $\hat{\Lambda}_{iCP}(\tau)=\sum_{j=1}^{J}d\hat{\Lambda}_{iCP}(t_j)$. 
Similarly, we define the \textit{average counterfactual hazard}, $\bar{\Lambda}_{iCP}(\tau)$, as the mean across the multiverse:
\begin{align*}
\bar{\Lambda}_{iCP}(\tau):=\frac{1}{J}\sum_{j=1}^{J}\frac{d_j}{m}. 
\end{align*}
Contrastingly, the risk in the actual world, denoted $F(\tau)$, sums the actual deaths $d\Ntid(t_j)$ over the sample size:
\begin{align*}
F(\tau):=\frac{\sum_{j=1}^Jd\Ntid(t_j)}{m}, 
\end{align*}
Under the assumption that deaths at time $t_j$ in the actual world correspond to deaths in the $j$-th possible world, $d\Ntid(t_j) \leq d_j$, with equality if all deaths occur simultaneously or at the first event time (i.e., $j=1$).

In any possible world within the multiverse, the number of deaths does not exceed the total deaths in the actual world, $d_j\leq \sum_{j=1}^Jd\Ntid(t_j)$. It follows that $\sum_{j=1}^{J}d_j\leq \sum_{j=1}^{J}\sum_{j=1}^{J}d\Ntid(t_j)=J\sum_{j=1}^{J}d\Ntid(t_j)$. 
Conversely, as previously explained, $d\Ntid(t_j)\leq d_j$. Combining these inequalities, we derive $J^{-1}\sum_{j=1}^{J}d_j\leq \sum_{j=1}^{J}d\Ntid(t_j)\leq \sum_{j=1}^{J}d_j$, leading to:
\begin{align*}
\bar{\Lambda}_{iCP}(\tau)\leq F(\tau)\leq \Lambda_{iCP}(\tau). 
\end{align*}
These inequalities indicate that the risk in the actual world lies between the average iCP hazard and the cumulative iCP hazard, serving as lower and upper bounds, respectively.

\section{Discussion}
\label{ss:discussion}

This paper has the following contributions. Firstly, we introduce the iCP-hazard, a counterfactual hazard expressed as a marginal probability via a counterfactual process. Unlike traditional approaches that condition on prior survival status, the iCP-hazard intervenes in this status instead. While the identification formula of the iCP-hazard, conditional on covariates, aligns with that derived from the cCT-hazard defined by counterfactual survival time, the two differ in their handling of covariates due to the non-collapsibility of the cCP-hazard. Additionally, compared to counterfactual survival time, the process-based hazard is more straightforward to illustrate in a causal diagram.
Secondly, we establish an underlying pseudo-population for the counterfactual iCP-hazard. Specifically, we conceptualize a multiverse expanded by the number of event times, where the risk in each possible world corresponds to the hazard at each time point.
Lastly, by linking hazards with risks in the multiverse, we enable causal inference that draws from developments in counterfactual risk, potentially addressing issues associated with hazard analysis.

We've demonstrated that the proposed iCP hazard functions as a controlled direct effect, with the prior outcome status acting as the mediator. Identification of this hazard relies on the assumption of no unmeasured confounding: $\Ntid(t; z, n(t^-))\indep (Z, \Ntid(t^-))\mid X$. As depicted in Figure~\ref{ff:dag}, adjustments must be made for treatment-outcome and outcome-outcome confounders. The latter may vary over time, as the confounders for $\Ntid(t_1)$ and $\Ntid(t_2)$ may differ from those for $\Ntid(t_3)$ and $\Ntid(t_4)$. In practice, identifying and accounting for all time-dependent confounders is challenging. Even with available data, proper adjustment for time-varying confounding is essential to ensure a valid causal interpretation \citep{msm1, msm2}.
When the assumption of no unmeasured confounding is not met, alternative approaches may be employed to partially identify the iCP hazard. For instance, if monotonicity exists between the unmeasured confounder and the treatment, mediator, and outcome, bounds have been established for the controlled direct effect \citep{vanderweele2011controlled}. On the other hand, if we have no information about the unmeasured confounder, one may also use two instruments, one for the treatment, and the other for the mediator, to identify the controlled direct effect among compliers \citep{frolich2017direct}.

Time holds a crucial position in discussions of causality. Traditionally, we've regarded time as a background coordinate, akin to physical dimensions, to pinpoint causal relationships among objects, assuming that causation follows the temporal flow~\citep{sep-causation-metaphysics}.
However, when time itself becomes a random variable in causal models, it becomes challenging to conceptualize time as a causal variable within the temporal coordinate. Utilizing a counterfactual process, rather than counterfactual survival time, to describe survival status offers the advantage of preserving time's role as a coordinate.
When we intervene in a time-dependent variable, we not only create a possible world (as with interventions in time-independent variables) but also specify a distinct time coordinate from that of the actual world. This creates an independent spatiotemporal space, enabling agents to gather data and conduct statistical analyses.

There are both similarities and differences between the classical possible world semantics for counterfactual conditionals and the proposed multiverse semantics for counterfactual hazards.
Similarities lie in the creation of alternative worlds for comparison in causal inference and the artificial nature of these created worlds. While one might argue that the possible worlds of hazards, where individuals die at specific time points, diverge from reality, it's not uncommon for classical possible world semantics to employ artificial settings (particularly in discussions of epistemology, such as the notion of being a brain in a vat~\citep{sep-epistemology,sep-modality-epistemology}).
The primary difference lies in the variability of world settings and the procedure for evaluating counterfactual conditionals or hazards. Classical possible world semantics aim to encompass every conceivable scenario, resulting in worlds that vary in numerous aspects. In contrast, possible worlds of hazard in the proposed multiverse semantics only vary in one dimension—specifically, the time point(s) at which individuals can die.
Regarding procedure, the truth value of counterfactual conditionals is determined by comparing the actual world with a selected set of worlds that closely resemble the actual world in the classical possible world framework. In the multiverse semantics, however, all possible worlds are combined into a multiverse, and the comparison involves two multiverses --- one under $z=0$ and another under $z=1$ --- to ascertain the causal effect.

The multiverse semantics offers a preferable interpretation of the iCP hazard for two main reasons.
Firstly, in each possible world, the agent behaves identically to the agent in the actual world. This eliminates the need for the agent to traverse multiple worlds to piece together fragments of hazards, as required in the direct interpretation.
Secondly, both the counterfactual hazard and counterfactual risk are consistent across all possible worlds, ensuring a well-defined pseudo-population behind the cumulative hazard. Consequently, unlike the actual risk, which only offers a causal interpretation for a period starting from $t_0$, the counterfactual hazard (or risk) can provide a valid causal interpretation for any arbitrary period of time.


\newpage
\bibliographystyle{myplainnat}
\bibliography{counthaz}






\end{document}